\documentclass[sigconf]{acmart}

\acmYear{2026}
\copyrightyear{2026}
\acmConference[FSE Companion '26]{34th ACM Joint European Software Engineering Conference and Symposium on the Foundations of Software Engineering}{July 5--9, 2026}{Montreal, QC, Canada}
\acmBooktitle{34th ACM Joint European Software Engineering Conference and Symposium on the Foundations of Software Engineering, July 5--9, 2026, Montreal, QC, Canada}
\acmPrice{}
\acmDOI{}
\acmISBN{}

\usepackage{float}
\usepackage{placeins}
\usepackage{pgfplots}
\usepackage{tikz}
\usepackage{xcolor}
\usepackage{enumitem}

\begin{document}

\title{Mapping GitHub Sponsorships: A Longitudinal Observatory for Open-Source Sustainability}

\author{Rylan Hiltz}
\affiliation{
  \department{Department of Computer Science}
  \institution{Trent University}
  \city{Peterborough}
  \state{Ontario}
  \country{Canada}
}
\email{rylanhiltz@trentu.ca}

\author{Taher A. Ghaleb}
\orcid{0000-0001-9336-7298}
\affiliation{
  \department{Department of Computer Science}
  \institution{Trent University}
  \city{Peterborough}
  \state{Ontario}
  \country{Canada}
}
\email{taherghaleb@trentu.ca}

\begin{abstract}
\vspace{-1pt}
Financial sustainability is vital for open-source software, yet systematic research on funding remains limited. GitHub Sponsors, launched in 2019 as a direct developer-to-developer funding model, lacks bulk API access, hindering large-scale studies. This paper introduces a live, continuously operating observatory for tracking and analyzing the GitHub Sponsors ecosystem. The observatory performs priority-based graph traversal with daily incremental updates, real-time normalization, and exposes collected data through an interactive dashboard and analysis-ready CSV exports. A sample dataset collected during a 72-hour run captures 49K+ users across 144 countries and serves as an example of the tool's output, not a fixed deliverable. An interactive dashboard (\url{https://github-sponsorships.com}) enables practitioners and researchers to explore sponsorship patterns, filter by geography and demographics, and benchmark against funded peers. Preliminary results on the sample show strong participation asymmetries and geographic concentration, suggesting several research directions.
\end{abstract}


\keywords{GitHub sponsors, Open-source sustainability, Software funding}

\maketitle

\section{Introduction}
Open-source sustainability relies on diverse funding models, but empirical understanding of these mechanisms is still limited. Prior work has studied corporate sponsorship~\cite{Wen2024,Medappa2023}, foundation grants~\cite{Overney2020}, and dual licensing~\cite{Comino2011,Valimaki2003}, yet direct developer-to-developer funding remains underexplored. GitHub Sponsors, which supported over 49,148 developers as of March 2026, represents a major funding channel where individual developers and organizations provide direct financial support to open-source contributors.

GitHub’s APIs expose extensive repository data but lack bulk sponsorship endpoints. This has limited empirical research: prior work~\cite{Shimada2022} analyzed 1.2M repositories yet found only 9.3K sponsored developers, missing most sponsors. Without standardized data collection, researchers cannot study how sponsorship affects project outcomes, developer behavior, or long-term software sustainability.

This paper introduces a live observatory tool for continuously tracking and analyzing the GitHub Sponsors ecosystem. The tool performs priority-based graph traversal, daily incremental updates, and real-time normalization, and exposes collected data through an interactive dashboard and analysis-ready CSV exports. The dataset is an artifact produced by the running tool, not a one-time snapshot.

\vspace{1.2pt}
\noindent Our contributions are:
\vspace{-3pt}
\begin{enumerate}[leftmargin=2em]
    \item \textbf{Live/Open Source Observatory.} A continuously operating, open-source tool performing priority-based graph traversal, automatic rate limiting, daily incremental updates, and standardized normalization to maintain fresh longitudinal dataset.

    \item \textbf{Interactive Dashboard.} A publicly accessible dashboard (\url{https://github-sponsorships.com}) that enables practitioners and researchers to filter, sort, and explore sponsorship patterns, compare profiles with funded peers, and identify traits linked to funding success.

    \item \textbf{Sample Dataset.} A 72-hour sample collected by the observatory, capturing 49K+ users across 144 countries, enriched with demographic, geographic, and activity metadata and exportable as CSV. This is an example output of the running tool, not a static deliverable.
    
    \item \textbf{Empirical Preliminary Results.} Initial exploration on a sample extracted by the observatory highlights participation asymmetries and geographic concentration.

\end{enumerate}

\vspace{1.2pt}
\noindent\textbf{Paper Organization.}
Section~\ref{sec:research_landscape} positions this work within prior research.
Section~\ref{sec:implementation} describes the data collection infrastructure.
Section~\ref{sec:dataset} details dataset structure and access.
Section~\ref{sec:findings} presents preliminary empirical results.
Section~\ref{sec:use_cases} demonstrates research applications.
Section~\ref{sec:limitations} discusses limitations and threats to validity.
Section~\ref{sec:future} outlines future work.
Section~\ref{sec:conclusion} concludes the paper.

\vspace{-3pt}
\section{Research Landscape \& Motivation}
\label{sec:research_landscape}

\subsection{Open-Source Sustainability Research}
Software sustainability research has established that financial models significantly impact project longevity~\cite{McGuire2023,venters2023sustainable}, code quality, and community health~\cite{Overney2020}. However, most studies focus on traditional funding: corporate employment~\cite{Medappa2023}, foundation grants~\cite{Wen2024}, or dual licensing~\cite{Comino2011}. Direct peer-to-peer funding platforms like GitHub Sponsors, Patreon, and Open Collective represent a newer funding model whose effects on software development remain understudied.

\vspace{-3pt}
\subsection{The GitHub Sponsors Gap}
\vspace{-1pt}
GitHub Sponsors, launched in May 2019, enables developers to receive monthly or one-time payments. Sponsorable developers create a profile listing support tiers (e.g., \$5/month, \$50/month), each with optional perks. Sponsors can also make one-time contributions; these appear on the recipient's profile temporarily but are not reflected in the public sponsor count after the transaction completes, making one-time support systematically undercounted in any dataset relying on public profile data. As of March 2026, it supports over 49,148 sponsorable users---a 232\% increase from 14,892 in June 2022~\cite{Conti2023}. Despite this growth, systematic empirical research remains limited by data access constraints. Recent work has begun to examine the program's effects: Wang et al.~\cite{wang2022influence} found that sponsorship correlates with increased developer activity on GitHub, and Fan et al.~\cite{fan2024my} showed that social media mentions on Twitter/X can drive sponsorship growth. Our observatory enables larger-scale replication and extension of such studies.

GitHub provides APIs for repository metrics but no bulk endpoint for sponsorship data. Researchers must query individual profiles, facing strict rate limits, or sample projects, missing most participants. Existing infrastructures like GHTorrent~\cite{gousios2012ghtorrent} do not capture sponsorships and are not updated regularly, while the GraphQL API requires per-user queries, making large-scale collection infeasible.

\vspace{-2pt}
\section{Observatory Infrastructure}
\label{sec:implementation}
The observatory was designed to address several key challenges: ensuring broad coverage beyond just high-profile projects, keeping data fresh with continuous updates, standardizing messy profile details, and making the process reproducible and extensible.

\vspace{-2pt}
\subsection{Architecture}
\vspace{-1pt}
The core of the observatory is a back-end worker designed to run continuously with minimal downtime. It adopts a layered architecture separating data collection, normalization, storage, and export:
\vspace{-2pt}
\begin{itemize}[leftmargin=1.85em]
    \item \textbf{API/Controller Layer (\texttt{/api}).} Implemented as a Flask Blueprint, it handles HTTP routes for data export, user filtering, and statistics. For example, \texttt{/api/users} validates and filters queries from the frontend, while the \texttt{/api/export} endpoint generates CSV files with customizable fields for research use.

    \item \textbf{Business Logic Layer (\texttt{/ingest}).} The \texttt{IngestWorker} implements priority-based graph traversal, managing the discovery queue and coordinating data collection. Activity collection is performed per-year per-user, requiring separate API calls for each year of an account's history, a resource-intensive step the worker schedules intelligently to avoid redundant calls.

    \item \textbf{Data Access Layer (\texttt{/db/queries}).} Abstracts SQL operations for user records, sponsorship edges, and enrichment metadata.
    
    \item \textbf{Utilities Layer (\texttt{/utils}).} Provides GitHub API wrappers with automatic rate limiting and retry logic, plus normalization services (location geocoding, demographic inference).
\end{itemize}

The frontend is a single-page application built with React, TailwindCSS, and the Ant Design component library, providing a responsive and interactive interface with dynamic dashboards, sortable tables, and visualizations of sponsorship, activity, and demographic data. The structured CSV export provides an analysis-ready research artifact generated by the running tool, supporting integration with R, Python, and other analysis tools.

\vspace{-2pt}
\subsection{Graph Traversal}
\begin{figure}[ht!]
  \centering
  \vspace{-4pt}
  \includegraphics[width=.9\linewidth]{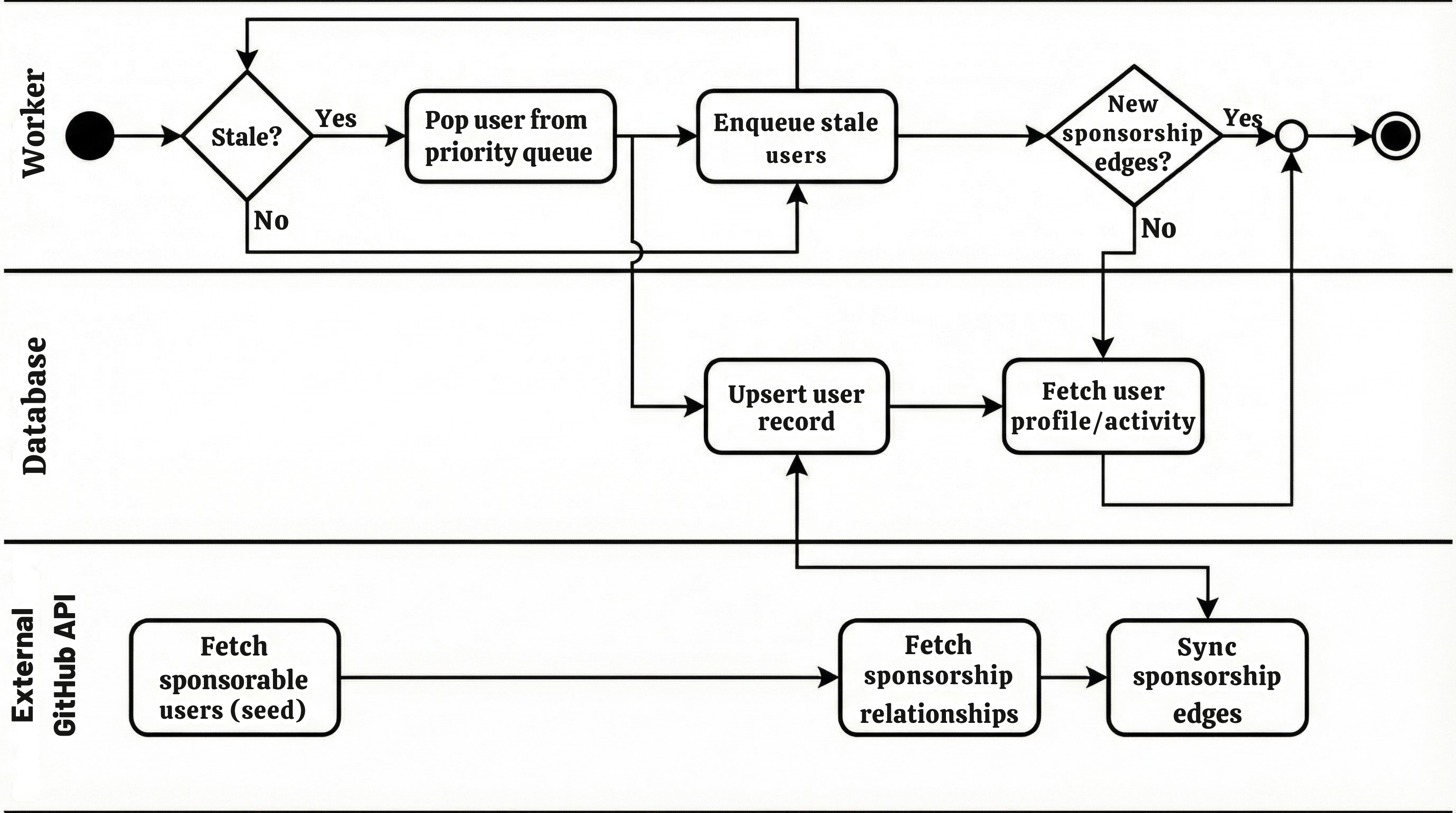}
  \vspace{-8pt}
  \caption{Data collection workflow showing the \texttt{IngestWorker}, database, and GitHub API interactions.}
  \Description{Activity Diagram with three swimlanes showing the worker's decision points, database operations, and API calls in the data collection process.}
  \label{Fig:GitHub-Activity-Diagram}
  \vspace{-6pt}
\end{figure}

\vspace{-1pt}
Figure~\ref{Fig:GitHub-Activity-Diagram} illustrates the workflow of the longitudinal observatory collector. The observatory traverses the GitHub Sponsors network using a queue, starting from \texttt{Sponsorable} users and iteratively exploring sponsorship relationships. Users are dequeued and checked for staleness; stale users are re-enqueued. For each user, the system checks database presence, fetches profile and activity data if needed, and updates the database. New sponsorship edges are synchronized, and newly discovered users are enqueued. Active users are prioritized, while disconnected users deprioritized to support longitudinal coverage. The architecture separates traversal logic, persistent storage, and API interactions in a loop where the database continuously feeds updated users back into the queue. Traversal covers all users connected via sponsorship edges from the initial seed; users reachable only through non-sponsorable accounts represent a negligible fraction of the ecosystem.

Key properties of the traversal include:
\vspace{-2pt}
\begin{itemize}[leftmargin=1.5em]
    \item \textbf{Seed Coverage.} The traversal initializes from all users marked \texttt{Sponsorable} by GitHub (49,148 users as of March 2026), ensuring broad coverage of the sponsorship ecosystem.
    
    \item \textbf{Priority-Based Exploration.} Users with active sponsorship relationships are refreshed more frequently than inactive users, enabling efficient longitudinal updates.
    
    \item \textbf{Discovery Completeness.} All users connected to the seed set via sponsorship edges are discovered; users linked exclusively to non-sponsorable accounts are rare and have negligible impact on coverage.
\end{itemize}

\vspace{-2pt}
\subsection{Data Normalization Protocols}
Raw GitHub data requires normalization for research use:
\vspace{-2pt}
\begin{itemize}[leftmargin=1.85em]
    \item \textbf{Location Standardization.} User-provided location strings (``NYC'', ``New York'', ``USA'') are geocoded via OpenStreetMap API, resolving to standardized country names. We retain the importance score to prioritize unambiguous matches.
    
    \item \textbf{Demographic Enrichment.} We extract explicitly specified pronouns from user profiles using headless browsing. For users without pronouns, we record this fact without inference, avoiding the methodological weaknesses.
    
    \item \textbf{Activity Metrics.} We track yearly contribution counts (commits, pull requests, issues, code reviews) to analyze how development activity correlates with sponsorship success.
\end{itemize}

\vspace{-3pt}
\subsection{Research-Grade Observatory Infrastructure}
The observatory provides reproducible, scalable, high-quality GitHub Sponsors data collection beyond ad-hoc methods.
\vspace{-2pt}
\begin{itemize}[leftmargin=1.55em]
    \item \textbf{Rate Limit Handling.} To collect 49K+ users and their sponsorship activity ($\approx$150,000 API calls) under GitHub’s 5,000 requests/hour limit, the observatory uses automatic backoff, token rotation, and request queuing, completing the process in 72 hours.
    
    \item \textbf{Incremental Updates.} Sponsorship relationships evolve daily. The priority queue ensures active users are refreshed more frequently than inactive ones, keeping collected data current while avoiding redundant API calls.
    
    \item \textbf{Data Quality.} Robust normalization protocols handle 23+ location-string variations and validate all geocoding results; without it, 38\% of records would have unusable location data.
    
    \item \textbf{Reproducibility.} Versioned snapshots with collection timestamps enable replication studies and longitudinal comparisons, unlike one-off snapshots generated by ad-hoc methods.
\end{itemize}

\vspace{-3pt}

\vspace{-2pt}
\subsection{Tool Maturity, Deployment \& Usability}
\vspace{-1pt}
Operational since September 2025, the observatory has recorded 63,549 sponsorships as of March 2026 and supports research on open-source sustainability. Visitor interactions are tracked via Google Analytics. The codebase is roughly evenly split between frontend and backend components, with automated tests covering both. A public instance is available at \url{https://github-sponsorships.com}, while the full source code can be deployed locally via Docker. The collector can be extended for specialized data enrichment, and the documentation provides setup instructions, API references, and example analysis notebooks for future research.

\section{Structure \& Access of Generated Data}
\label{sec:dataset}
\vspace{-1pt}
\subsection{CSV Export Schema}
\vspace{-1pt}
The observatory exports analysis-ready CSVs covering user identifiers, sponsorship metrics, demographics, activity, and metadata. All fields are documented with normalization procedures and quality indicators in the repository. Estimated earnings are computed as a lower bound: the developer's minimum published tier multiplied by their sponsor count. Actual earnings are higher when sponsors select premium tiers, but tier selection is not publicly available.

\vspace{-3pt}
\subsection{Data Quality Indicators}
\vspace{-2pt}
We provide explicit quality flags for each record, indicating confidence in demographic data to support informed analyses:

\vspace{-2pt}
\begin{itemize}[leftmargin=1.85em]
    \item \textbf{High Confidence.} User has pronouns AND a geocoded location with importance $> 0.8$. Suitable for demographic analysis.
    
    \item \textbf{Medium Confidence.} Either pronouns OR location data is reliable. Usable with appropriate caveats.
    
    \item \textbf{Low Confidence.} Missing or ambiguous data on both dimensions. May be excluded from demographic studies.
\end{itemize}

\vspace{-2pt}
\subsection{Researcher Access and Example Analyses}
Researchers can import the CSV directly into statistical tools. We provide example analyses for common research tasks, including sponsorship network centrality, logistic regression models predicting sponsorship success, and geographic disparity analysis. We will continue to expand these resources.

\vspace{-2pt}
\section{Preliminary Empirical Results}
\label{sec:findings}
We conducted a 72-hour collection run using the observatory to produce a sample dataset. This sample is not exhaustive; it represents the tool's state at one point in time and will grow with subsequent runs. The run produced 49K+ enriched users and organizations with active sponsorships.

\subsection{Participation Asymmetry}
Table~\ref{tab:sponsorship_activity} summarizes sponsorship participation by user type. Participation is highly asymmetric: 40{,}549 users sponsor others, while only 7{,}343 receive sponsorship, a 5.5:1 ratio. This indicates broad community support concentrated on a smaller set of recipients. A small subset of all users (1{,}256; 2.6\%) both sponsor and receive sponsorship, potentially acting as bridges within the sponsorship network.

\begin{table}[ht]
  \centering
  \vspace{-5pt}
  \renewcommand{\arraystretch}{0.77}
  \caption{Sponsorship Participation by User Type}
  \vspace{-10pt}
  \label{tab:sponsorship_activity}
  \begin{tabular}{lrrrr}
    \toprule
    \textbf{Type} & \textbf{Sponsored} & \textbf{Sponsoring} & \textbf{Both} & \textbf{Total} \\
    \midrule
    User & 6,077 & 38,121 & 1,172 & 45,304 \\
    Org & 1,266 & 2,428 & 84 & 3,728 \\
    \midrule
    \textbf{Total} & \textbf{7,343} & \textbf{40,549} & \textbf{1,256} & \textbf{49,148} \\
    \bottomrule
  \end{tabular}
  \vspace{-13pt}
\end{table}

\vspace{2pt}
\subsection{Demographic Coverage}
\vspace{-2pt}
We observe that, of 49,148 users, 6,128 (12\%) explicitly specify pronouns: 87\% use masculine pronouns, 9\% feminine, and 4\% other/neutral. The remaining 43,252 users (88\%) do not specify pronouns.
We do not infer pronouns for these users to avoid methodological bias. Future work could examine whether pronoun specification correlates with sponsorship, while accounting for selection effects.

\subsection{Geographic Concentration}
\vspace{-2pt}
Table~\ref{tab:location_role_distribution} summarizes sponsorship participation across the top six countries. The United States accounts for 16.6\% of all sponsorships (7,910 users), followed by Germany (6.2\%), the UK (3.8\%), Japan (2.7\%), Canada (2.2\%), and France (2.0\%). Together, these countries represent 33.4\% of all observed sponsorships.
Overall, the sample spans 144 countries, with geocoded locations available for 52.3\% of users.
This concentration suggests potential geographic inequities in sponsorship access, motivating further analysis that controls for regional and infrastructural confounders.

\begin{table}[ht!]
  \centering
  \vspace{-3pt}
  \renewcommand{\arraystretch}{0.77}
  \caption{Top 6 Countries by Sponsorship Participation}
  \vspace{-10pt}
  \label{tab:location_role_distribution}
  \begin{tabular}{lrrrr}
    \toprule
    \textbf{Country} & \textbf{Sponsored} & \textbf{Sponsoring} & \textbf{Both} & \textbf{Total} \\
    \midrule
    USA     & 1,320 & 6,292 & 298 & 7,910 \\
    Germany & 520   & 2,323 & 111 & 2,954 \\
    UK      & 450   & 1,270 & 74  & 1,794 \\
    Japan   & 294   & 879   & 100 & 1,273 \\
    Canada  & 206   & 804   & 58  & 1,068 \\
    France  & 299   & 580   & 50  & 929   \\
    \midrule
    \textbf{Total} & \textbf{3,089} & \textbf{12,148} & \textbf{691} & \textbf{15,928} \\
    \bottomrule
  \end{tabular}
\vspace{-12pt}
\end{table}

\section{Research/Practitioner Applications}
\label{sec:use_cases}

\vspace{-2pt}
\subsection{Example Research Designs}
\vspace{-1pt}
The observatory can be integrated with established resources to enable richer empirical studies.
\vspace{-2pt}
\begin{itemize}[leftmargin=1.85em]
    \item \textbf{Survival Analysis.} Joining with \textit{GHArchive} lets researchers link sponsored users to their repositories and model project abandonment with Cox proportional hazards models to test whether sponsorship correlates with longer maintenance.
    
   \item \textbf{Regression Analysis.} Integration with \textit{World of Code} enables logistic regression modeling of sponsorship probability using comprehensive cross-platform contributions, project prominence (stars, forks), and geographic location.
    
    \item \textbf{Network Analysis.} Integration with \textit{Stack Overflow} data allows researchers to compute centrality metrics (PageRank, betweenness) and examine how community engagement relates to structure and influence within the funding network.
\end{itemize}

\vspace{-4pt}
\subsection{Reproducibility \& Extension}
\vspace{-2pt}
The observatory supports versioned snapshot releases, enabling:
\vspace{-2pt}
\begin{itemize}[leftmargin=1.85em]
    \item \textbf{Longitudinal Analysis \& Replication.} Versioned snapshots enable time-series studies and validation of findings.

    \item \textbf{Targeted Recollection.} The observatory enables focused collection on specific subpopulations (e.g., Python developers, security researchers) for specialized analyses.
\end{itemize}

\vspace{-3pt}
\subsection{Practitioner Applications}
\vspace{-2pt}
Beyond supporting empirical research, the observatory’s dashboard enables evidence-based decision-making for developers seeking sponsorship. Practitioners can explore contribution patterns, project types, and geographic contexts associated with funding success, benchmark themselves against funded peers, and identify actionable strategies to improve sponsorship prospects.

\vspace{-3pt}
\section{Limitations \& Threats to Validity}
\label{sec:limitations}
\vspace{-2pt}
\noindent\textbf{Data Coverage.}
Our traversal architecture discovers users connected to the initial sponsorable seed set. Developers who sponsor others but are not themselves sponsorable may be missed. However, this represents a small fraction given that sponsoring requires creating a GitHub account, which makes one discoverable.

\vspace{2pt}
\noindent\textbf{Temporal Dynamics.}
The preliminary sample is a 72-hour collection run. Longitudinal claims require multiple collection runs over time. We plan quarterly releases to support time-series analysis.

\vspace{2pt}
\noindent\textbf{Demographic Limitations.}
Pronoun specification is voluntary and may correlate with other unobserved characteristics (cultural context, age, technical background). Analyses using demographic data must acknowledge selection bias. We deliberately avoid inferring pronouns to prevent compounding this limitation.

\vspace{2pt}
\noindent\textbf{Geographic Biases.}
Location normalization depends on OpenStreetMap data quality, which varies by region. Developers using VPNs, privacy-protecting location strings (``Remote'', ``Earth''), or ambiguous strings may be misclassified or excluded.

\vspace{2pt}
\noindent\noindent\textbf{Sponsorship/Activity Over Time.}
The dashboard displays auxiliary metrics like commit history alongside sponsorship data. Such contextual metrics help profile developers but are not directly linked to sponsorship outcomes. Given that GitHub lacks historical snapshots of sponsorship counts, it is not yet possible to overlay activity trends with sponsorship growth over time. We plan to address this limitation through quarterly versioned releases.

\vspace{-4.5pt}
\section{Future Work}
\label{sec:future}
\vspace{-2pt}
We plan to release quarterly snapshots of sponsorship profiles and activities to enable longitudinal analysis of sponsorship lifecycle events. Future enrichment will link users to core project metadata, detailed contribution patterns, and published sponsorship tier structures. We also plan ego-network visualization tools for exploring sponsorship neighborhoods around specific developers.

\vspace{2pt}
\noindent\textbf{Public Research Platform.}
We deployed a public web dashboard as an online community resource, offering interactive visualizations, filtering, and exploration, alongside CSV exports. This supports both practitioners and researchers in hypothesis generation and confirmatory statistical studies.

\vspace{2pt}
\noindent\textbf{Research Questions Enabled.}
The observatory enables the following research questions for future investigations.
\vspace{-2pt}
\begin{itemize}[leftmargin=1.85em]
    \item \textbf{RQ1: Sustainability Impact.}
    Is sponsorship associated with reduced project abandonment or altered maintenance patterns?
    
    \item \textbf{RQ2: Developer Incentives.}
    Which developer characteristics predict sponsorship, and how is effort distributed across sponsored and non-sponsored projects?
    
    \item \textbf{RQ3: Geographic Equity.}
    Are there geographic sponsorship biases, especially for developers in lower-income regions?
    
    \item \textbf{RQ4: Network Effects.}
    Are sponsorship networks centralized, and how do they overlap with code collaboration networks?
    
    \item \textbf{RQ5: Longitudinal Dynamics.}
    How do sponsorships change over time, and what events drive their growth or decline?
\end{itemize}
\vspace{-2pt}

\vspace{-4.5pt}
\section{Conclusion}
\label{sec:conclusion}
\vspace{-2.5pt}
This work presents a live observatory for systematic empirical investigation of GitHub Sponsors as a funding mechanism. The observatory performs priority-based graph traversal, applies standardized normalization, and generates analysis-ready exports for diverse research methods.
Preliminary results show strong participation asymmetries and geographic concentration, indicating key directions for further study. By releasing reproducible observatory infrastructure and versioned observatory snapshots, this work supports research on developer incentives, geographic equity, network effects, and funding dynamics, enabling evidence-based strategies for open-source project sustainability, offering practitioners actionable insights via the public dashboard.

\vspace{-4.5pt}
\section*{Availability of Artifacts}
The observatory infrastructure, sample dataset, example analysis Jupyter notebooks, and demonstration video are available at:

\vspace{2pt}
\noindent\textbf{Observatory Infrastructure \& Dataset:} \url{https://github.com/Taher-Ghaleb/Github-Sponsor-Dashboard}.

\vspace{2pt}
\noindent\textbf{Docker Deployment:} Full-stack deployment available via Docker Compose.

\vspace{2pt}
\noindent\textbf{Interactive Dashboard:} \url{https://github-sponsorships.com}.

\vspace{2pt}
\noindent\textbf{Demonstration Screencast} [\url{https://youtu.be/TYBfRpCUgEw}]:

\noindent(4-minute walkthrough of the deployed dashboard and repository).

\vspace{-1pt}
\begin{acks}
\vspace{-1pt}
This work is funded by the Natural Sciences and Engineering Research Council of Canada (NSERC): RGPIN-2025-05897.
\end{acks}

\balance
\bibliographystyle{ACM-Reference-Format}
\bibliography{main}

\end{document}